\documentclass[preprint,aps,prb,epsf]{revtex4}   
\include{amsmath}
\include{epsfx}
\usepackage{graphicx}
\begin{document}
\title{{\bf Hydrodynamic Interactions in Protein Folding}}
\author{{\bf Marek Cieplak and Szymon Niewieczerza{\l}}}

\address{
Institute of Physics, Polish Academy of Sciences,
Al. Lotnik\'ow 32/46, 02-668 Warsaw, Poland\\}

\begin{abstract}
We incorporate hydrodynamic interactions (HI) in a coarse-grained and
structure-based model of proteins by employing the Rotne-Prager
hydrodynamic tensor. We study several small proteins and demonstrate
that HI facilitate folding. We also study  HIV-1 protease
and show that HI make the flap closing dynamics faster. The HI are found
to affect time correlation functions in the vicinity of the native state
even though they have no impact on same time characteristics 
of the structure fluctuations around the native state.
\end{abstract}
\maketitle

\section{Introduction}

Globular proteins acquire distinct compact native conformations in
water as a result of the hydrophobic effect.
Another role of water is to mediate
the hydrodynamic interactions (HI) between moving amino acids
in analogy to the HI in polymers \cite{Kirkwood,Zimm} and
between particles in colloidal suspensions
\cite{Mazur,Brady,Felderhof,Ladd,Cichocki,Dhont}.
One expects that the HI should generally enhance cooperative
features in the dynamics of proteins, but it is not clear in
what way, exactly, will this show.
In the case of mechanically-induced unfolding, the HI have been found
to lead to a) reduction in peak unfolding forces when stretching
at high steady velocities \cite{Szymczak1}, b) reduction in unfolding times
when stretching at constant force \cite{Szymczak1} because of the dragging
effect, and c) hindering of unravelling imposed through uniform \cite{Szymczak2} 
and shear \cite{Szymczak3} fluid flows because of the screening effects.

In this paper, we focus on the kinetics
of folding and of fluctuational motions around the native state.
We assess the relevance of HI in these phenomena
within the previously used coarse-grained implicit-solvent model
of proteins \cite{Hoang3,Hoang,Sulkowska_1,Sulkowska_2} of $N$ beads with
the Lennard-Jones contact interactions. As in refs.
\cite{Szymczak1,Szymczak2}, the HI are introduced through the
Rotne-Prager tensor \cite{Rotne,Yamakawa}.

Kikuchi et al. \cite{Kikuchi} have taken another approach to introduce
the fluid-related effects. Instead of employing the tensorial field they make use
of the stochastic rotation dynamics \cite{Malevanets}.
They have demonstrated that HI facilitate
the collapse transition of a self-attracting homopolymer because of 
dragging effect in which
a bead attracted to another bead through, say, the Lennard-Jones potential
drags the fluid containing other beads with it. 
They have also, together with a
related thesis work of Ryder\cite{Ryder}, reported a small,
of order 10\%, reduction  in the folding time, $t_{fold}$, in simple models
of several proteins, such as 2ci2. 
On the other hand, Baumketner and Hiwatari \cite{Baumketner} claim otherwise,
pointing out that HI give rise to the effective repulsion between two beads 
which are coming towards each other, thus slowing down the collapse.
They consider a simplified "minimal" model introduced in ref. \cite{Honeycutt}.
For a short $\beta$-hairpin, they obtain a certain increase in the folding 
time and lack of any effect for a short $\alpha$-helix.

Here, we consider four short proteins, 1l2y ($N$=20), 1bba ($N$=36),
1crn ($N$=46) 2ci2 ($N$=65) and one $\beta$-hairpin ($N$=14) 
-- a fragment of the 1gb1 protein.
In each case, our results are consistent with the picture of
Kikuchi et al. \cite{Kikuchi}. 
The collapse phase of the folding process is indeed faster.
However, we find the reduction in $t_{fold}$
to be more substantial than in the work of Ryder \cite{Ryder}, 
for instance, by a factor, $f$, of 2 for 1crn.
Possible reasons for this quantitative discrepancy include differences in the
shape of the initial conformations (Ryder may have more compact starting 
conformations compared to the fully extended situations considered here) 
and in the precise nature of the folding criterion (Ryder may have a criterion
based on a characteristic size of a distance deviation from the native state
as opposed to the contact-based criterion we use).
Our results differ qualitatively with that of ref. \cite{Baumketner}. It should 
noted that the model used there comes with attractive contacts only between 
hydrophobic amino acids and requires this attraction to be the strongest at the 
distance of 4.2 {\AA} between the corresponding C$^{\alpha}$'s.
In proteins, contact distances between the C$^{\alpha}$ atoms
extend in fact from 4.4 to 12.8 {\AA} \cite{Hoang2} which is likely to change the
nature of the effects associated with the HI.
In fact, when we a) constrain the potentials for the $beta$-hairpin case 
to have a minimum at 4.2 {\AA} the reduction factor goes down to 1.2 and then, in
addition, b) increase the number of allowed contacts to imitate the
situation envisioned within the minimal model, $f$ becomes equal to 0.87,
indicating that the HI make the folding longer. Thus the result of
ref. \cite{Baumketner} is related to the minimal nature of the model considered
there.

We then consider a larger protein -- the HIV-1 protease (1a30). This protein
is a homodimer and each of the two chains comprises 98 amino acids.
Its active site is covered by two flexible $\beta$-hairpins, called flaps, 
that controll the entry of a polypeptide substrate. The flap opening dynamics
in this protein has been studied before within all atom \cite{Harte}
and coarse-grained molecular dynamics schemes. Here, we pull the flaps 
apart and then monitor their return to the native form as a function of time.
We observe that the HI make the flap closing faster.

Finally, we return to the small protein 1l2y and
investigate fluctuations around the native state by considering
single- and double-residue characteristics of these fluctuations. We find
that the HI have no impact on their same-time averages. 
This result is not surprising because, in the overdamped limit
considered here, the equilibrium distributions of conformations are the same.
However, it sets the stage for the observation that time 
correlation functions are sensitive to the HI despite the protein
being essentialy folded.

\section{Methods}

The coarse-grained, Go-type model \cite{Goabe} of a protein we use is constructed
based on the knowledge of the native state. There are many ways to implement it.
Examples are given in refs. \cite{Honeycutt,Veitshans,Nymeyer,Clementi,Hoang0,models}.
The details of the particular implementation we use are described in 
\cite{Sulkowska_1,models}.
Each residue is represented by a single bead centered on the
position of the C$^{\alpha}$ atom. The successive beads along the backbone are
tethered by harmonic potentials with a minimum at 3.8 {\AA} and they are
also endowed with the chirality based local backbone stiffness \cite{Sulkowska_1}.
The other interactions
between the residues $i$ and $j$ are split into two classes: native and non-native
as described in ref. \cite{Sulkowska_1}.
The native contacts are endowed with the effective Lennard-Jones potential
$V_{ij} = 4\epsilon \left[ \left( \frac{\sigma_{ij}}{r_{ij}}
\right)^{12}-\left(\frac{\sigma_{ij}}{r_{ij}}\right)^6\right]$
whereas the non-native contacts are purely repulsive.
The length parameters $\sigma _{ij}$ are chosen so that the potential
minima correspond, pair-by-pair, to the experimentally established
native distances between the respective aminoacids.
The energy parameter, $\epsilon$,
is taken to be uniform and 0.3$\epsilon /k_B$ is usually playing the
role of the room temperature \cite{Sulkowska_2}.

The time evolution is described by the Brownian dynamics (BD) \cite{ermak}. 
The displacement of particle $i$ in time step $\Delta t$ obeys
\begin{equation}
 {\bf r}_i  - {\bf r}_i^0 =  \sum_j \bigl( \nabla_j \cdot {\bf D}_{ij}^0 \bigr) \Delta t
+ \frac{1}{k_B T} \sum_j {\bf D}_{ij}^o \cdot {\bf F}^{0}_j \Delta t + {\bf B}_i,
\label{nar1}
\end{equation}
where the index $0$ denotes the values of respective quantities at the beginning of the
time step, ${\bf F}_j$ is the force exerted on particle $j$ by
other particles, and $T$ is temperature.
${\bf D}$ is a diffusion tensor and  ${\bf B}$ - a random displacement given
by a Gaussian distribution with an average value of zero and covariance
obeying
 \begin{equation}
< {\bf B_i} {\bf B}_j > = 2 {\bf D}^0_{ij}
\Delta t.
\label{Gauss}
\end{equation}
If the diffusion tensor is nondiagonal, there exists 
a hydrodynamic  coupling between particles $i$ and $j$
({\it cf.} Eq.\ref{nar1}).
We take the diffusion tensor in the form \cite{Rotne,Yamakawa}
\begin{equation}
{\bf D}_{ii}=\frac{k_B T}{\gamma} {\bf I}
\end{equation}
and
\begin{equation}
{\bf D}_{ij}=\displaystyle \frac{k_B T}{\gamma} \frac{3a}{4 r_{ij}} \left\{\begin{array}{cl}
      \displaystyle \left[ \left( 1+\frac{2a^2}{3 r_{ij}^2} \right) {\bf I} + \left(1-
\frac
{2a^2}{r_{ij}^2}\right)
{\bf \hat{r}}_{ij} {\bf \hat{r}}_{ij} \right], & r_{ij} \geq 2a\ \\
\\
\displaystyle \frac{r_{ij}}{2a} \left[ \left( \frac{8}{3}-\frac{3r_{ij}}{4a} \right) {\bf I}
+
\frac
{r_{ij}}{4a}
{\bf \hat{r}}_{ij} {\bf \hat{r}}_{ij} \right], & r_{ij}  < 2a
         \end{array}\right.
\label{rp}
\end{equation}
where ${\bf r}_{ij}={\bf r}_{j}-{\bf r}_{i}$, $a$ is the hydrodynamic radius,
and $\gamma$ is the damping constant.
By Stokes's law, $\gamma=6 \pi a \eta$, where $\eta$ is the viscosity of the fluid.
The characteristic time scale in the problem, $\tau$,
is of order 1 ns, which reflects duration of a diffusional passage
of a typical contact distance ($\sim 5$ {\AA}).

The situation corresponding to the lack of HI will be denoted by BD (for Brownian
dynamics) and will be implemented by setting ${\bf D}_{ij}$ ($i\ne j$) to zero.
The selection of the value of the hydrodynamic radius is not obvious and
more studies are needed to settle this issue. We just want to determine
qualitatively what would happen if it was non-zero.
When one thinks of amino acids as represented by spheres then one would expect
that $a$ should not exceed a half of the distance between consecutive C$^{\alpha}$'s.
Thus we consider the hydrodynamic radius of 1.5 {\AA} to have $a$ substantial enough and 
yet satisfying this criterion. On the other hand,
the side groups extend laterally and may produce a correspondingly
larger drag force. It has been argued \cite{Torre,Antosiewicz} that a
characteristic $a$ of an amino acid could be even as big as 4.2 {\AA} whereas a
characteristic van der Waals radius is about 3.0 {\AA} \cite{Zamyatin}.
The van der Waals volume does not include hydratation layers.
Such big values would mean existence of an overlap between
spheres in a chain, but its usage takes into account the
non-spherical properties of the amino acids in an approximate way.
Considering $a$ of 1.5 {\AA} has a numerical advantage because the time unit
is governed by the damping constant and is thus proportional to $a$. Thus the
folding times are expected to scale linearly with $a$ for systems both with
and without HI, making it relevant to assess the role of HI regardless
the particular choice of $a$. It cannot be ruled out, however, that HI
may introduce some corrections to scaling. Assessment of such corrections 
would need a separate study.

It should be noted that there are some amino acid to amino acid variations
in the values of the van der Waals and the de la Torre - Bloomfield \cite{Torre}
hydrodynamic radii.
For glycine, alanine, and arginine, the former are 2.4 {\AA}, 2.6 {\AA}, 
and 3.3 {\AA} respectively, and the latter are 3.6 {\AA}, 3.7 {\AA}, 
and 4.5 {\AA} respectively \cite{Antosiewicz1}.
These variations  are not expected to affect
the findings in analogy to the lack of sensitivity to the variations
in the values of the masses \cite{Hoang} or in the values of the
damping constant \cite{Juelich}. A scaling by the mean value of the
parameters has been demonstrated in these other cases.

Throughout this paper we employ the Brownian dynamics evolution algorithm
since it allows for a straightforward incorporation of the HI.
However, all of the BD results obtained here are reproduced by the Langevin
dynamics algorithm as used in refs. \cite{Hoang,Sulkowska_1}.
The Langevin dynamics involves inertia but the system is overdamped.

\section{Results}

\subsection{Kinetics of folding}

Figure 1 shows the $T$-dependence of the median value of $t_{fold}$ determined
as the first time to establish all native contacts ($r_{ij} < 1.5 \sigma _{ij}$).
For each of the proteins studied, there is a
broad plateau of optimal folding. 
In the optimal regime, $t_{fold}$ with HI
is shorter by a factor $f$  than in the BD case. The value of the factor
depends on the protein and is indicated on the right hand
side of the figure. For the $\beta$-hairpin, $f$ is 1.3.
The largest protein studied, 2ci2, has been simulated only at one temperature
(0.3 $\epsilon /k_B$). The simulations 
yielded $t_{fold}$ of $\sim 840 \tau$ and $\sim 470 \tau$
for the BD and HI models respectively. The corresponding $f$ factor of 1.8
is clearly distinct from $\sim$1.1 found in ref.\cite{Ryder}.
The discrepancy may probably be attributed  to two circumstances:
a) the statistics  involved in ref. \cite{Ryder} have been restricted to several trajectories,
b) the values of $f$ depend on the nature of the starting conformations.
The bigger the number of the native contacts that are present in the initial
stages of folding, the smaller the value of $f$.
In our simulations, the starting conformations are fully extended.

The HI-induced acceleration of folding is governed mostly by the initial
collapse as illustrated for two typical trajectories for 1crn in the bottom 
panel of Figure 2. This figure shows the time evolution of the radius of
gyration, $R_g$. Qualitatively, the collapsed phase is said to be reached
when $R_g$ does not exceed 15\% of its native value.
In agreement with ref. \cite{Kikuchi}, the HI are seen to significantly
accelerate arrival at the collapsed phase (by a factor larger, $\sim 2.6$,
than the acceleration factor found for $t_{fold}$). The subsequent
establishment of contacts involves a stochastic search in the conformational
space and is not affected by any hydrodynamic effects.

The top panel of Figure 2 shows the distributions of the values of $t_{fold}$ 
across 100 trajectories for 1crn.
The distributions are non-Gaussian and have pronounced tails.
The peak of the distribution corresponding to HI is at shorter times than BD,
but the long-time ends are comparable.
It should be noted that both at high and low temperature ends, when the
folding time becomes large, the distinction between the HI and BD cases
evaporates because when the protein kinetics are slow the related
induced fluid flows are sluggish.

Figure 3 shows the scenario diagram \cite{Sulkowska_1} which represents the
first times needed to establish specific native contacts as averaged over
the trajectories. The native
contacts are labelled by their sequential separation $|j-i|$. It is seen that
the presence of HI accelerates establishment of all contacts without changing
the order in which they are set. The same statement applies to the
other proteins we have studied and we expect it to hold in general.

\subsection{Closing of flaps in HIV-1 protease}

The native structure of HIV-1 protease is schematically represented by the
lower right conformation shown in Figure 4. The hairpin fragments that form
the flaps are highlighted by showing the corresponding C$^{\alpha}$ atoms.
The locations of the flaps can be conveniently described by providing
coordinates of the four-atom centers of mass. 
The two centers of mass are $R_{f,nat}$=4.09 {\AA} away from each other.
We pull the model flaps apart
by exerting a stretching force applied along the direction that links the
initial centers of mass of the two flaps until they are separated by
32.75 {\AA}. The resulting conformation is shown on the upper left in
Figure 4. The force is then removed and we monitor the distance, $R_f$,
between the centers of mass as a function of time.

Figure 4 demonstrates that the HI make the closing of the flaps faster.
The value of $R_f=2R_{f,nat}$ is reached about twice as fast with HI
compared to the BD case. The phenomenon is analogous to refolding discussed
previously. However, it affects only a part of the full protein.

\subsection{Structure fluctuations around the native state}

We now consider 1l2y which is the smallest among the set of proteins
studied here so that an appropriate conformational statistics can be generated.
Its native structure is shown in Figure 5. It is akin to a $\beta$-hairpin
in wich one of the "arms", however, is shaped into an $\alpha$-helix.
The 20 amino acids are enumerated from the helical end.
We set the protein in its native conformation,
then evolve it at $T$=0.3$\epsilon /k_B$ in 
four different trajectories which last for 200 000 $\tau$.

\subsubsection{Equal time characteristics}

One way to quantify the dynamics of the motion around the native state
is to study the fluctuations of amino acid positions $<\Delta R_i^2(t)>$
around the average value. We follow the procedure of Kabsch and Sanders \cite{Kabsch}
and project any evolved conformation onto a reference conformation so that
the results are not affected by translation of the center of mass and by
rotation of the whole molecule. $<\Delta R_i^2>$ is defined then as 
$<\vec{r}_i^2> -<\vec{r}_i>^2$. The top panel of Figure 6 shows the 
modulations of $<\Delta R_i^2>$ along the sequence. The largest fluctuations
take place at the termini and at $i$=12 where the helix joins the other
arm of the hairpin. Smaller, but still substantial, fluctuations occur at
$i$= 8 and 15. The data points corresponding to the HI and BD cases nearly
overlap indicating the lack of relevance of the HI in the equilibrium fluctuations.

A similar statement applies to the 
two-point (equal time) characteristics of the motion
as illustrated in Figure 7 for pairs of residues 
$i$ and $j$ that make a native contact. There are 27
such contacts in 1l2y (counted by the index $l$)
and 13 of them are indicated by the lines in Figure 5.
The allocation of the index $l$ to its corresponding pair of $i$ and $j$
is also written underneath the data point in the top panel of Figure 7.

The simplest characteristic is $f_l = (<r_{ij}^2> - <r_{ij}>^2)^{1/2}$.
This quantity does not depend on any rigid body motion of the protein
and it rapidly acquires lack of dependence on the duration of the
measurement. The top panel of Figure 7 shows that it is 
insensitive to the inclusion of the HI.
The modulations in the strenghts of $f_l$ span
a factor of two. Large values indicate
high amplitude oscillations (like in contact 22 between amino acids
7 and 11) and small values point to a relative rigidness (like in contact
27 between 11 and 14). A removal of the "keystone" contact 18 
(between residues 6 and 17) is found to reduce $f_{22}$
and $f_{15}$ but to unhance $f_5$ and $f_9$. For an isolated $\alpha$-helix
the $f_l$s are usually small and uniform except for enhancements at the termini.
The $\beta$-hairpin has largest fluctuations at contacts that link the termini
($f_l \sim 1.5$ {\AA}).

Another popular characteristic is the so called dynamical cross-correlation map
(see e.g. refs. \cite{Harte,Chillemi,Giansanti}) defined as
\begin{equation}
C_l\;=\;C_{ij}\;=\; <\Delta {\bf r}_i \cdot \Delta {\bf r}_j> /
[<\Delta {\bf r}^2_i> \; < \Delta {\bf r}^2_j>]^{1/2} \;\;.
\end{equation}
Unlike $f_l$, $C_l$ involves features which are primarily orientational
and does not depend much on the distance between $i$ and $j$.
$C_l$ can be either positive or negative, depending on the
nature of the relative displacement. Its determination requires
making the rigid-body transformation that brings
the conformation closest to the reference structure. We find that
$C_l$ for 1l2y settles in their stationary values within 
2000 $\tau$. These stationary
values are shown in the bottom panel of Figure 7. The values of $C_l$ are
modulated but the sensitivity to the presence of HI is again seen to be minor.

All of the above single- and double-residue equal time characteristics
correlation functions may be also calculated
by using the Gaussian network model \cite{Bahar} and by performing
the normal mode analysis. 
We find that the nature of the sequence- and pair-dependent
modulations coming from this approach is as in Figures 6 and 7 respectively.


\subsubsection{Time-dependent characteristics}

We now take various conformations along each trajectory and investigate what 
happens to them a time $t$ later. The motion of the center of mass of the
protein is subtracted and we could get reasonable statistics for 
time intervals not exceeding 5000 $\tau$. We define the correlation function
\begin{equation}
S_i(t)= <\vec{r}_i(0)\cdot\vec{r}_i(t)>/<\vec{r}_i(0)\cdot\vec{r}_i(0)>
\end{equation}
and $S(t) = <S_i>_{av}$, where $<...>$ denotes an average over the
trajectories and $<...>_{av}$ an average over the residues.
The definition implies that for $t$=0, $S_i$=1.
The initial decay of $S(t)$ with $t$ is shown in Figure 8.
It is seen that the decay, or a 'decoherence', from the initial state 
is faster, when the HI are included. 
Due to limitations in statistics at longer delay times,
we cannot pinpoint the precise functional law for the time dependence of
$S(t)$. However, the important observations is that the time dependence
of time correlations is affected by the presence of the HI.

It should be noted here that there exists a
puzzle regarding the origin of long-time tails in the correlation
functions in proteins. Namely, the experiments by Xie and coworkers \cite{Xie2005}
seem to show that fluctuations in proteins decay very slowly in time,  
following a power law. A number of theoretical attempts to explain
this time-dependence have been made
\cite{Granek2005,Tang2006a,Tang2006b,Xing2006} but most of them predict the 
decay on a much shorter time-scales than those measured in experiments, or
perhaps they do not take into account the presence of cross links in the protein.


A related time-dependent characteristic is provided by
\begin{equation}
\Delta _i(t)= <[\vec{r}_i(t) - \vec{r}_i(0)]^2>/<<[\vec{r}_i(t) - \vec{r}_i(0)]^2>>_{av}
\end{equation}
in which the numerator would be related to the 
single-residue diffusion coefficient, $D_i(t)$, if divided
by $6t$. However, this $D_i(t)$ decays to zero, instead of homing in on a
constant, since the motion of the center of mass is excluded in the trajectory.
By dividing the numerator by the denominator in the expression for $\Delta _i(t)$,
one effectively removes most of the time dependence, whether it is provided by
BD or by HI. The interesting observation is, as demonstrated in the bottom
panel of Figure 6, that $\Delta _i(t)$ shows no difference
between BD and HI and, practically, has no time dependence.
Its sequential behavior is qualitatively similar to that of $<\Delta R_i^2>$
(the top panel of Figure 6).

In summary, the HI affect the time scale  of folding significantly by
making the collapse faster through the cooperative action of the drag forces.
This phenomenon can be equivalently described
by invoking a system with no HI but with a reduced effective viscosity
by about a factor of two. 
Thus in the context of protein folding, or phenomena similar to
the flap closing in HIV-1 protease, the presence of HI just shifts the 
effective value of the damping constant $\gamma$. 
Observing phenomena that can be clearly attributed to the HI
may then be difficult experimentally unless one investigates a broad range of
temperatures. In an equilibrium evolution, the HI
affect time correlation functions
without influencing equal time correlations such as the rms fluctuations
in the contact length.

This paper has benefited from many fruitful discussions
with P. Szymczak who also provided us with his numerical code.
We also appreciate discussions with A. Giansanti and M. Wojciechowski.
This work has been supported by the grant N N202 0852 33 from the Ministry
of Science and Higher Education in Poland.



\begin{thebibliography}{199}

\bibitem{Kirkwood}
J. G. Kirkwood and J. Riseman,
{\it J. Chem. Phys.}  {\bf 16}, 565-573 (1948).

\bibitem{Zimm}
B. H. Zimm,
{\it J. Chem. Phys.}  {\bf 24}, 269-278 (1956).

\bibitem{Mazur}
P. Mazur and W. van Saarlos,
{\it Physica A.} {\bf 115}, 21-57 (1982).

\bibitem{Brady}
L. Durlofsky, J. F. Brady, and G. Bossis,
{\it J. Fluid Mech.} {\bf 180}, 21-49 (1987).

\bibitem{Felderhof}
B. U. Felderhof,
{\it Physica A.} {\bf 151}, 1-16 (1988).

\bibitem{Ladd}
A. J. C. Ladd,
{\it  J. Chem. Phys.} {\bf 88}, 5051-5063 (1988).

\bibitem{Cichocki}
B. Cichocki, B. U. Felderhof, K. Hinsen, E. Wajnryb, and J. Blawzdziewicz,
{\it J. Chem. Phys.} {\bf 100}, 3780-3790 (1994).

\bibitem{Dhont}
J. K. G.  Dhont, {\it An Introduction to Dynamics of Colloids}. Elsevier, Amsterdam,
(1996).

\bibitem{Szymczak1}
 P. Szymczak and M. Cieplak,
{\it J. Phys.: Condens. Matter}. {\bf 19}, 258224 (2007).

\bibitem{Szymczak2}
P. Szymczak and M. Cieplak,
{\it J. Chem. Phys.} {\bf 125}, 164903 (2006).


\bibitem{Szymczak3}
P. Szymczak and M. Cieplak,
{\it J. Chem. Phys.} {\bf 127}, 155106 (2007).

\bibitem{Hoang3}
M. Cieplak, T. X. Hoang and M. O. Robbins,
{\it Proteins: Struct. Funct. Bio.} {\bf 49}, 114-124 (2002).

\bibitem{Hoang}
M. Cieplak, T. X. Hoang and M. O. Robbins,
{\it Proteins: Struct. Funct. Bio.} {\bf 56}, 285-297 (2004).

\bibitem{Sulkowska_1}
J. I. Su{\l}kowska and M. Cieplak,
{\it J. Phys.: Cond. Mat.} {\bf 19}, 283201 (2007)

\bibitem{Sulkowska_2}
J. I. Su{\l}kowska and M. Cieplak,
{\it Bioph. J.} {\bf 94}, 6-13 (2008).

\bibitem{Rotne}
J. Rotne and S. Prager,
{\it J. Chem. Phys.} {\bf 50}, 4831-4837 (1969).

\bibitem{Yamakawa}
H Yamakawa,
{\it J. Chem. Phys.} {\bf 53}, 436-443 (1970).

\bibitem{Kikuchi}
N. Kikuchi, J. F. Ryder, C. M. Pooley, and J. M. Yeomans,
{\it Phys. Rev. E} {\bf 71}, 061804 (2005).

\bibitem{Ryder}
J. F. Ryder,
{\it Mesoscopic simulations of complex fluids},
Ph.D. thesis, the University of Oxford (2005).

\bibitem{Baumketner}
A. Baumketner and Y. Hiwatari,
{\it J. Phys. Soc. Jap.} {\bf 71}, 3069-3079 (2002).

\bibitem{Honeycutt}
J. D. Honeycutt and D. Thirumalai,
{\it Biopolymers} {\bf 32}, 695-709 (1992).

\bibitem{Malevanets}
A. Malevantes and R. Kapral,
{\it J. Chem. Phys.} {\bf 110}, 8605-8613 (1999).

\bibitem{Hoang2}
M. Cieplak and T. X. Hoang,
{\it Biophys. J.} {\bf 84}, 475-488 (2003)

\bibitem{Harte}
W. E. Harte Jr., S. Swaminathan, and D. L. Beveridge,
{\it Proteins: Struct. Funct. Gen.} {\bf 13}, 175-194 (1992).

\bibitem{Tozzini}
V. Tozzini, J. Trylska, C.Chang, and J. A. McCammon,
{\it J. Struct. Biol.} {\bf 157}, 606-615 (2007).

\bibitem{Trylska}
J. Trylska, V. Tozzini, C. Chang, and J. A. McCammon,
{\it Bioph. J.} 92:4179-4187 (2007).

\bibitem{Dickinson}
E. Dickinson,
{\it Chem. Soc. Rev.} {\bf 14}, 421-455 (1985).

\bibitem{Goabe}
H. Abe and N. Go,
{\it Biopolymers} {\bf 20}, 1013-1031 (1981).
S. Takada,
{\it Proc. Natl. Acad. Sci. USA}. {\bf 96}, 11698-11700 (1999).

\bibitem{Veitshans}
T. Veitshans, D. Klimov, and D. Thirumalai,
{\it Folding Des.} {\bf 2}, 1-22 (1997).

\bibitem{Nymeyer}
H. Nymeyer, A. E. Garcia, and J. N. Onuchic,
{\it Proc. Natl. Acad. Sci. (USA)} {\bf 95}, 5921-5928 (1998).

\bibitem{Clementi}
C. Clementi, H. Nymeyer, and J. N. Onuchic,
{\it J. Mol. Biol.} {\bf 298}, 937-953 (2000).

\bibitem{Hoang0}
T. X. Hoang and M. Cieplak,
{\it J. Chem. Phys.} {\bf 112},  6851-6862 (2000).

\bibitem{models}
J. I. Su{\l}kowska and M. Cieplak,
{\it Biophys. J.} {\bf 95}, 3174-3191  (2008).

\bibitem{ermak}
D. L. Ermak and J. A. McCammon,
{\it J. Chem. Phys.}, {\bf 69}, 1352-1360 (1978).

\bibitem{Torre}
J. Garcia de la Torre and V. A. Bloomfield,
{\it Quarter. Rev. Biophys.} {\bf 14}, 81-139 (1981).

\bibitem{Antosiewicz}
J. Antosiewicz and D. Porschke,
{\it J. Phys. Chem.} {\bf 93}, 5301-5305 (1989).

\bibitem{Zamyatin}
A. A. Zamyatin, 
Prog. Biophys. Mol. Biol. {\bf 24}, 107-123 (1972).

\bibitem{Antosiewicz1}
J. Antosiewicz -- private communication.

\bibitem{Juelich}
P. Szymczak and M. Cieplak,
NIC Workshop 2007 From Computational Biophysics to Systems Biology 2007,
ed. U. H. E. Hansmann, J. Meinke, S. Mohanty, and O. Zimmerman,
NIC Series Volume 36, 1-7 (2007).


\bibitem{Kabsch}
W. Kabsch and C. Sander,
{\it Biopolymers.} {\bf 22}, 2577-2637 (1983).

\bibitem{Chillemi}
G. Chillemi, P. Fiorani, P. Benedetti, and A. Desideri,
{\it Nucl. Acid. Res.} {\bf 31}, 1525-1535 (2003).

\bibitem{Giansanti}
F. Pizzitutti, A. Giansanti, P. Ballario, P. Ornaghi, P. Torreri,
G. Ciccotti, and P. Filetici,
{\it J. Mol. Recognit.} {\bf 19}, 1-9 (2006).

\bibitem{Bahar}
I. Bahar, A. R. Atilgan, and B. Erman,
{\it Folding and Design.} {\bf 2}, 173-181 (1997).

\bibitem{Xie2005}
W. Min, G. B. Luo, B. J. Cherayil, S. C. Kou, and X. S. Xie,
{\it Phys. Rev. lett.} {\bf 94}, 198302 (2005).

\bibitem{Granek2005}
R. Granek and J. Klafter,
{\it Phys. Rev. Lett.} {\bf 95}, 098106 (2005). 

\bibitem{Tang2006a}
J. Tang and R. A. Marcus,
{\it Phys. Rev. E.} {\bf 73}, 022102 (2006).

\bibitem{Tang2006b}
J. Tang and S. H. Lin,
{\it Phys. Rev. E.} {\bf 73}, 061108 (2006).

\bibitem{Xing2006}
J. H. Xing and K. S. Kim,
{\it Phys. Rev. E.} {\bf 74}, 061911 (2006).

\end{thebibliography}

\newpage

\centerline{FIGURE CAPTIONS}

\begin{description}

\item[Fig. 1. ]
The median folding times as a function of temperature for the three
proteins. The dotted lines and open data points correspond to the
BD model whereas the solid lines and data ponts to the model with the
HI. The hydrodynamic radius is 1.5 {\AA}. The $f$-factors on the right
give the ratios of the optimal  folding times without the HI to those
with the HI.

\item[Fig. 2. ]
Top panel: The distribution of single trajectory folding times at $k_BT/\epsilon$=0.3
for the model crambin. Bottom panel: Examples of time evolution of the
radius of gyration. The down-pointing arrows indicate entries into the
phase of compact collapsed conformations. The up-pointing arrows show
arrivals at the native conformation.

\item[Fig. 3. ]
The scenario diagram for folding of the model 1crn. The hexagons correspond
to the time to establish the constacts between C and I for the first time.
The circles correspond to the contacts between two $\beta$-strands.
The full circles - to the contacts within helices and between helices.
The asterisks correspond to all other contacts.
The upper symbols involve the HI and the lower do not.

\item{Fig. 4}
The flap dynamics in 1a30 as discussed in the main text. The conformation
on the right is native and on the left is stretched.
The dotted line corresponds to the BD case and is based on 50 trajectories.
The solid line corresponds to the HI case and is based on 20 trajectories,
all starting from the same stretched state.

\item[Fig. 5. ]
The native state of 1l2y. The larger numerical symbols enumerate the
amino acids from the N terminal. The smaller symbols enumerate native contacts
(the C-terminal generates no native contacts). The contacts are also indicated
by lines. The solid lines correspond to contacts with strong distance fluctuations,
as shown in Figure 6, whereas the dashed lines to those with weak fluctuations.

\item[Fig. 6. ]
Top panel: Variance in single-residue position fluctuations  as enumerated
sequentially. The solid (open) symbols are
for the model with (without) the HI and $k_BT/\epsilon$=0.3.
Bottom panel: normalized time-dependent variance $\Delta _i(t)$ for $t=1000\tau$.
The symbols corresponding to HI and BD in this and following two figures are
displaced laterally around the proper position to enhance their individual visibility.

\item[Fig. 7. ]
Fluctuations in the native contacts around the native state
at $k_BT/\epsilon$=0.3.
Top panel: the distance rms fluctuations.
The numbers indicate pairs of amino acids that are are connected by
the native contact labeled by $l$.
Bottom panel: the orientational rms fluctuations (known also as
the dynamical cross-correlations).

\item[Fig. 8.]
The normalized time correlation $S(t)$ as a function of time.

\end{description}

\begin{figure}
\epsfxsize=7.6in
\centerline{\epsffile{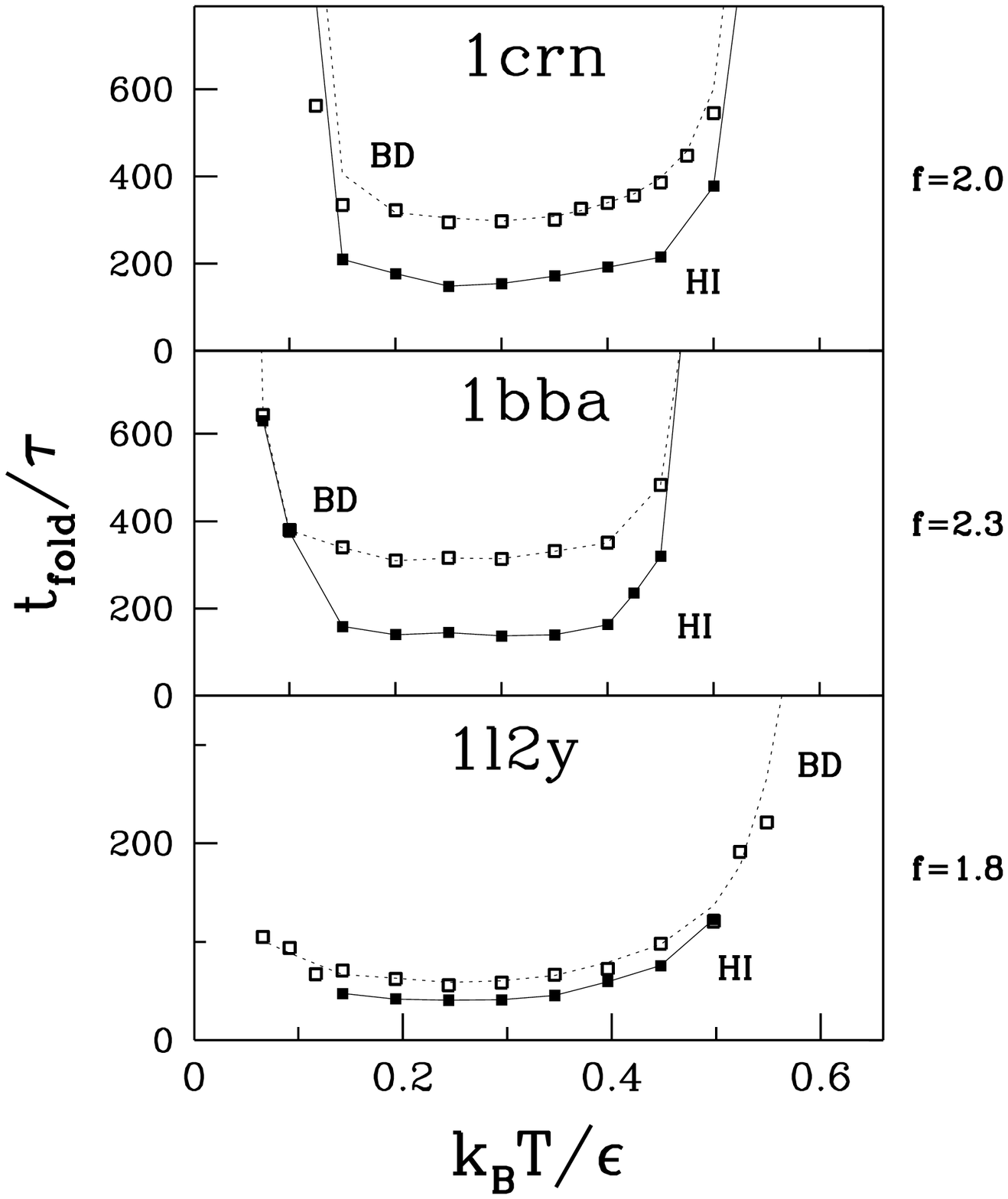}}
\caption{ }
\end{figure}

\vspace*{-4cm}


\begin{figure}
\epsfxsize=7.6in
\centerline{\epsffile{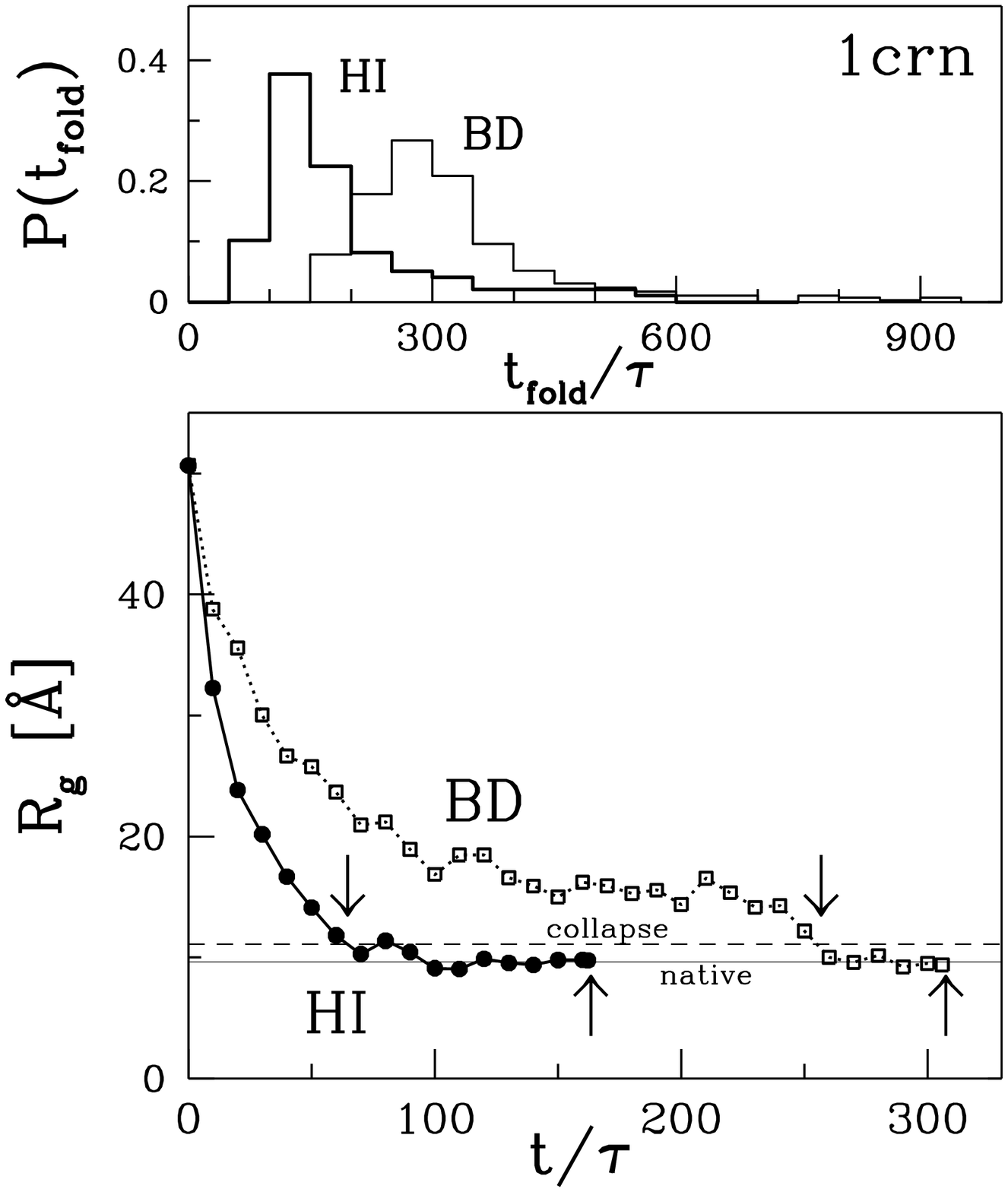}}
\caption{ }
\end{figure}

\vspace*{-4cm}

\begin{figure}
\epsfxsize=7.6in
\centerline{\epsffile{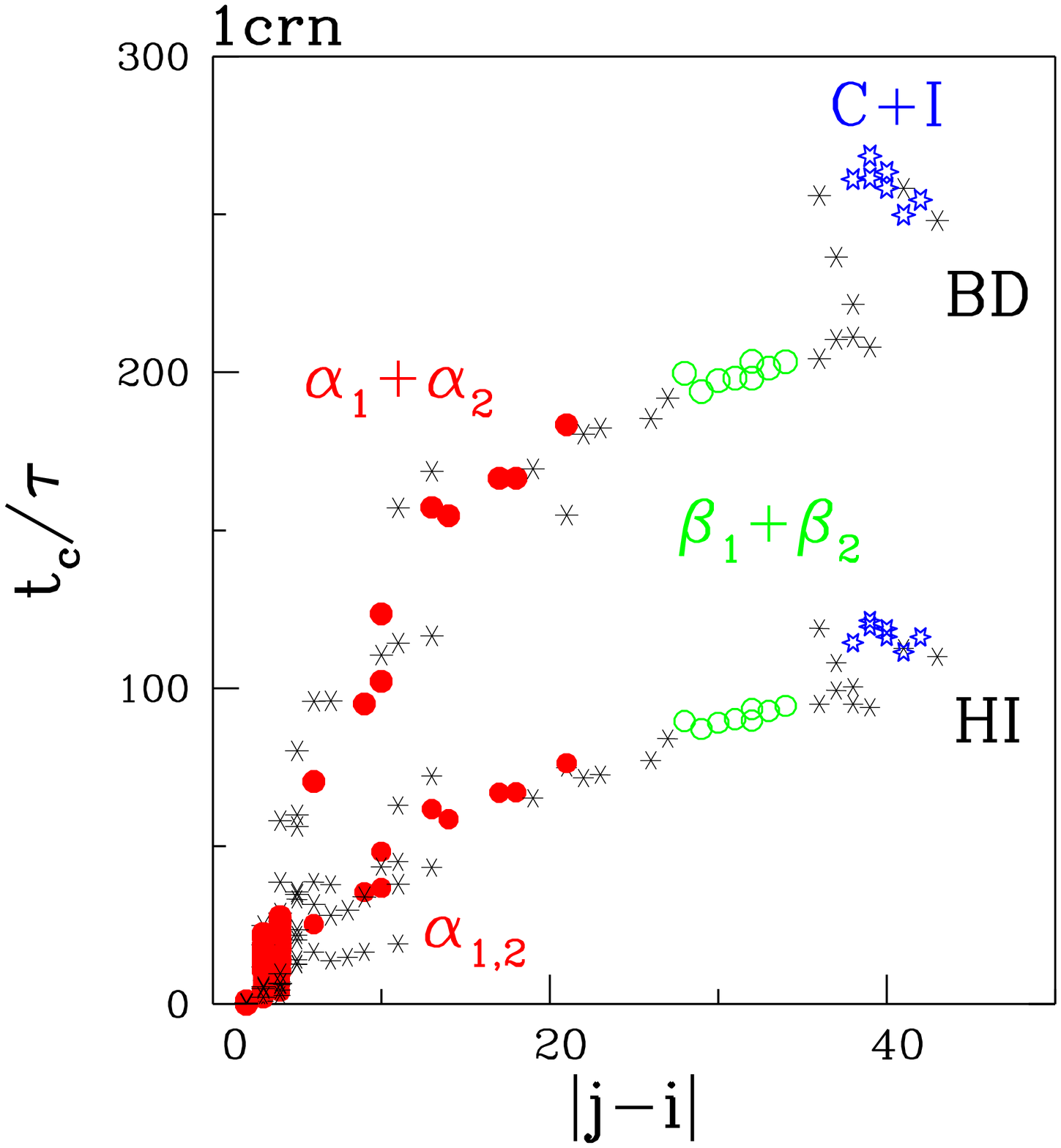}}
\caption{ }
\end{figure}

\vspace*{-4cm}

\begin{figure}
\epsfxsize=7.2in
\centerline{\epsffile{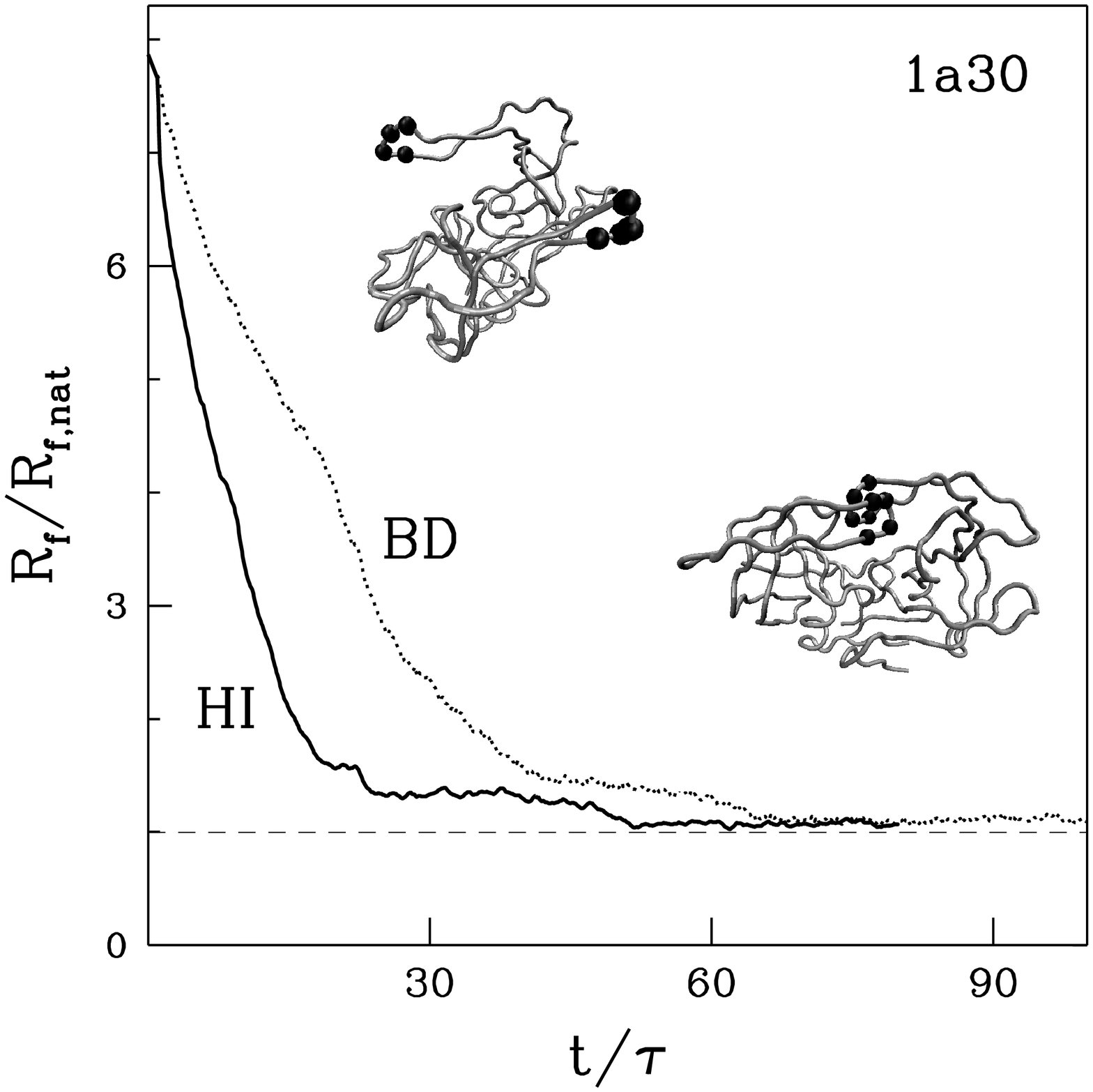}}
\vspace*{-1.9cm}
\caption{ }
\end{figure}

\vspace*{-4cm}

\begin{figure}
\epsfxsize=7.6in
\centerline{\epsffile{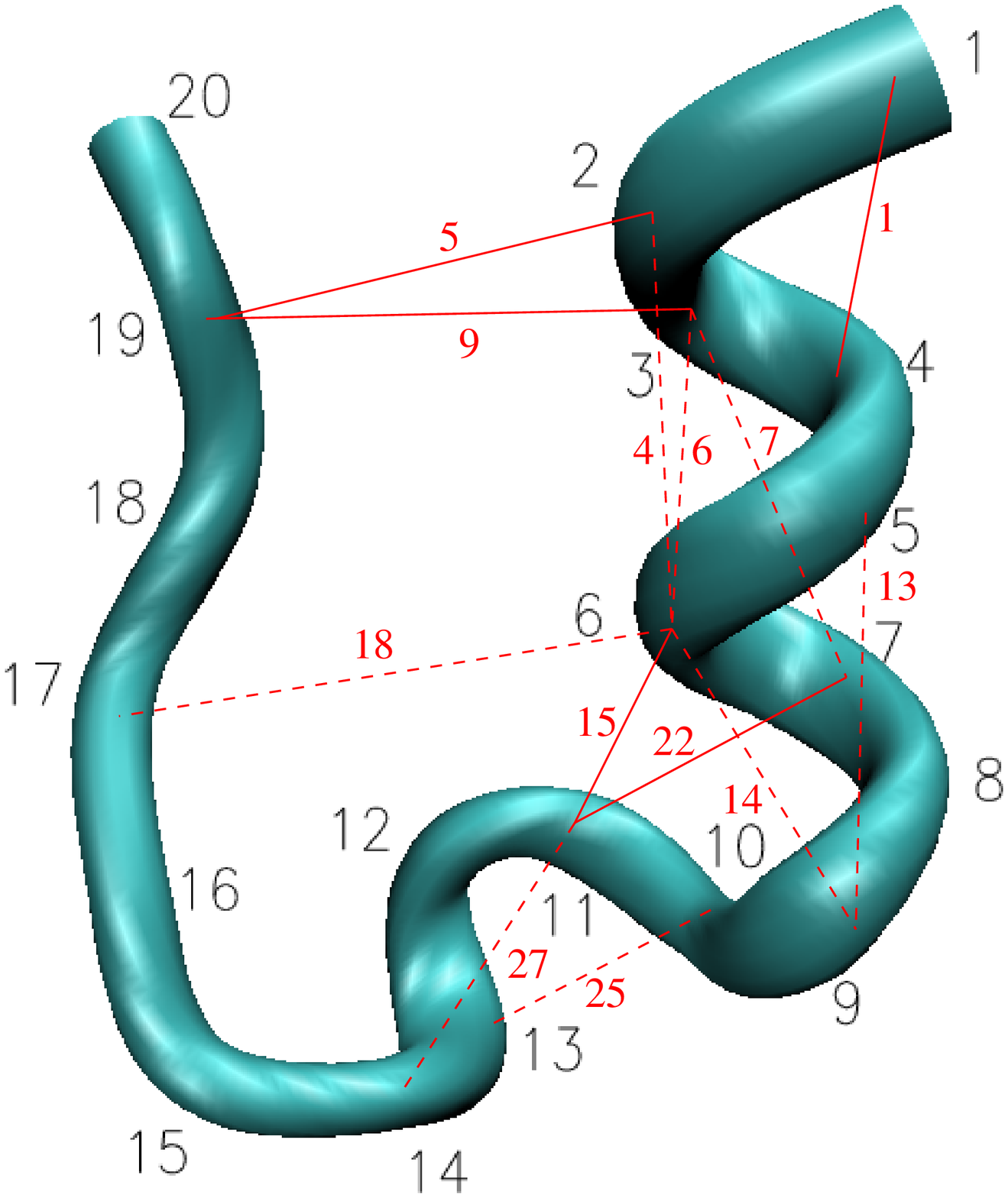}}
\caption{ }
\end{figure}

\vspace*{-4cm}

\begin{figure}
\epsfxsize=7.6in
\centerline{\epsffile{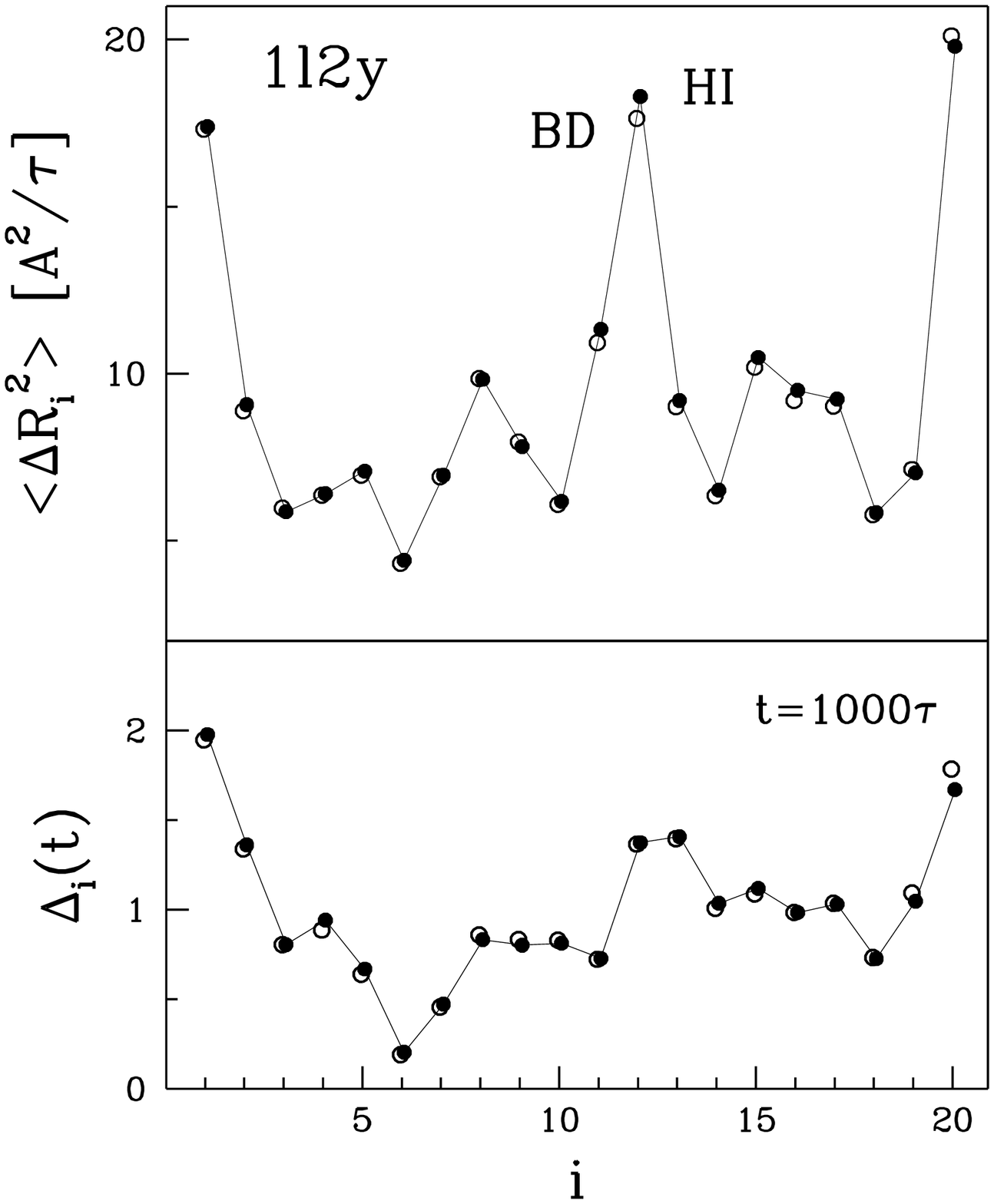}}
\caption{ }
\end{figure}

\vspace*{-4cm}

\begin{figure}
\epsfxsize=7.6in
\centerline{\epsffile{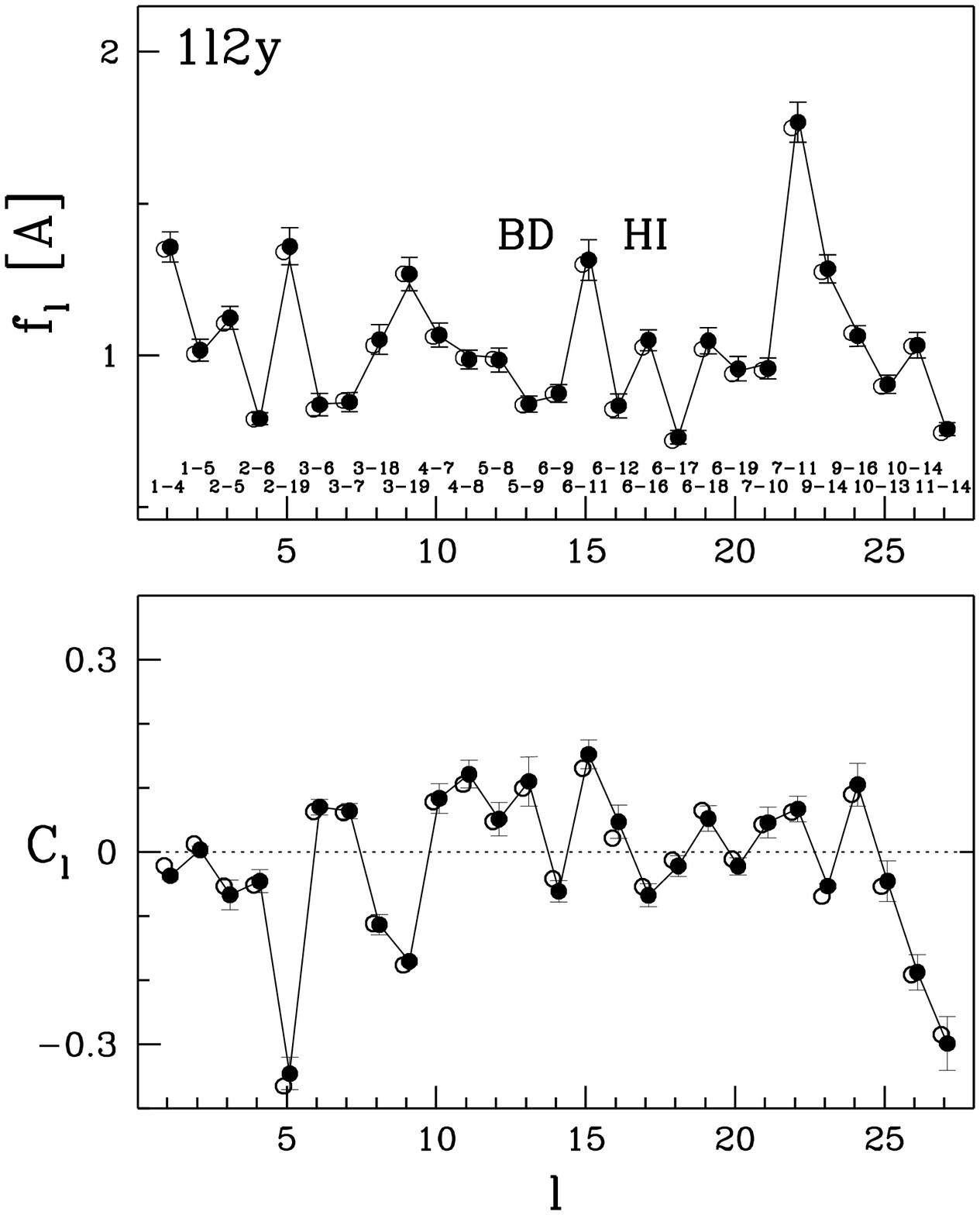}}
\caption{ }
\end{figure}

\vspace*{-4cm}

\begin{figure}
\epsfxsize=7.6in
\centerline{\epsffile{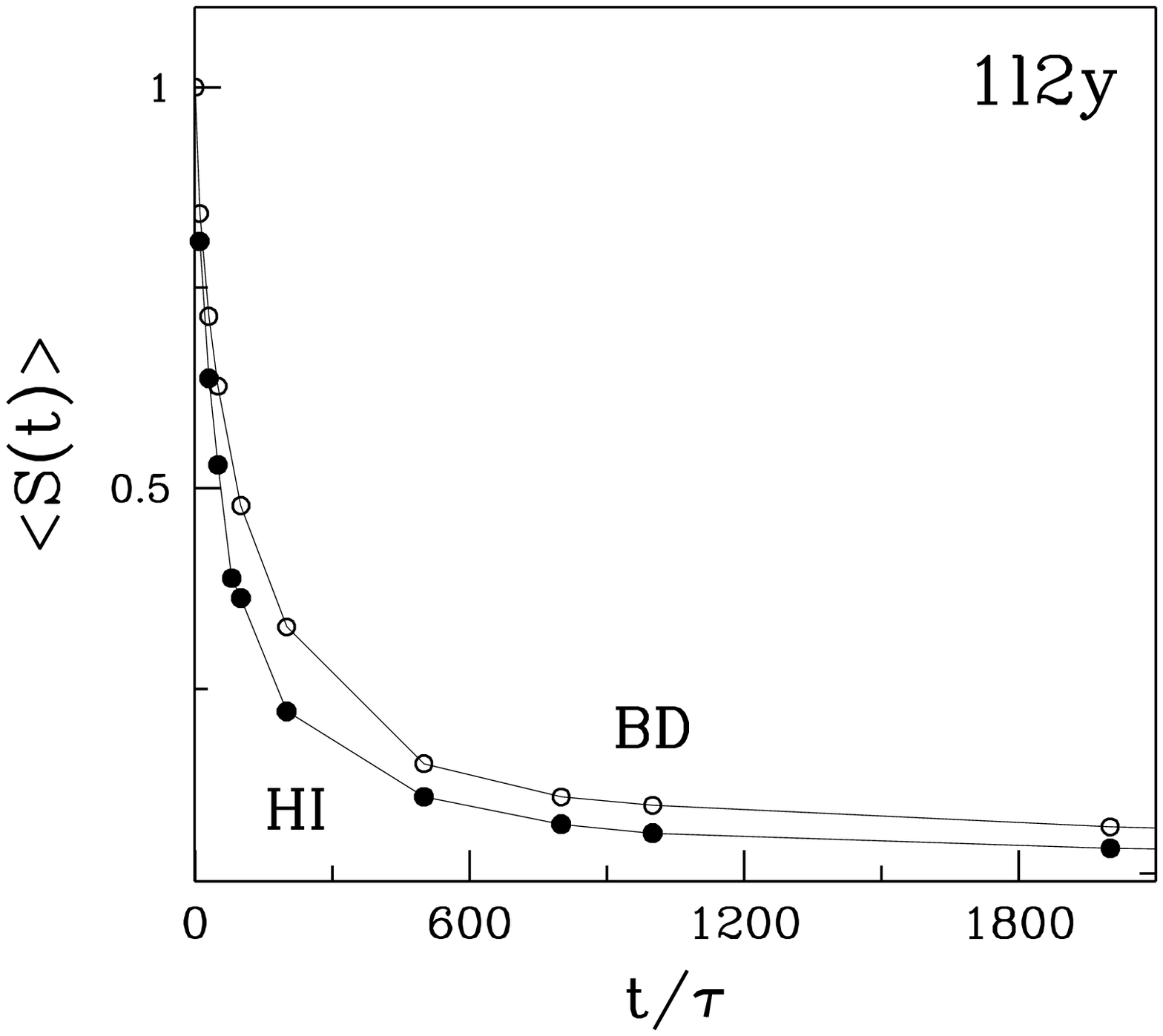}}
\caption{ }
\end{figure}

\vspace*{-4cm}

\end{document}